\documentclass[prb,twoside,twocolumn,10pt,floatfix,showpacs,citeautoscript,superscriptaddress]{revtex4-2}
\usepackage{amsmath}
\usepackage{amssymb}
\usepackage{booktabs}
\usepackage{braket}
\usepackage{float}
\usepackage{glossaries}
\usepackage{graphicx}
\usepackage{tabularx}
\usepackage{siunitx}
\usepackage{xspace}
\usepackage{xcolor}
\usepackage[version=4]{mhchem}
\usepackage[colorlinks=true,linkcolor=black,citecolor=black,filecolor=black,urlcolor=black]{hyperref}
\usepackage[T1]{fontenc} 

\DeclareSIUnit{\angstrom}{\textup{\AA}}

\graphicspath{{./figures/}}

\renewcommand{\vec}[1]{\ensuremath\boldsymbol{#1}}
\newcommand{\gpaw}{GPAW}
\newcommand{\deltakick}{\texorpdfstring{\ensuremath{\delta}}{delta}-kick}
\newcommand{\dd}{\mathrm{d}}

\setacronymstyle{long-short}
\newacronym{gllbsc}{GLLB-sc}{Gritsenko-van Leeuwen-van Lenthe-Baerends-solid-correlation}
\newacronym{hc}{HC}{hot carrier}
\newacronym{ks}{KS}{Kohn-Sham}
\newacronym{lcao}{LCAO}{linear combination of atomic orbitals}
\newacronym{ld}{LD}{Landau damping}
\newacronym{lsp}{LSP}{localized surface plasmon}
\newacronym{np}{NP}{nanoparticle}
\newacronym{rttddft}{RT-TDDFT}{real-time time-dependent density functional theory}
\newacronym[longplural={projected densities of state}]{pdos}{PDOS}{projected DOS}
\newacronym{xc}{XC}{exchange correlation}

\newcommand{\phys}{
    Department of Physics,
    Chalmers University of Technology,
    SE-412~96 Gothenburg, Sweden
}
\newcommand{\aalto}{
    Department of Applied Physics,
    Aalto University,
    FI-00076 Aalto, Finland
}
\newcommand{\unsw}{
    School of Chemical Engineering,
    The University of New South Wales,
    2052 Sydney, NSW, Australia
}

\begin{document}

\title{
    Tailoring hot-carrier distributions of \texorpdfstring{\\}{}plasmonic nanostructures through surface alloying
}
\author{Jakub Fojt}
\affiliation{\phys}
\author{Tuomas P. Rossi}
\affiliation{\aalto}
\author{Priyank V. Kumar}
\email{priyank.kumar@unsw.edu.au}
\affiliation{\unsw}
\author{Paul Erhart}
\email{erhart@chalmers.se}
\affiliation{\phys}

\begin{abstract}
Alloyed metal nanoparticles are a promising platform for plasmonically enabled hot-carrier generation, which can be used to drive photochemical reactions.
Although the non-plasmonic component in these systems has been investigated for its potential to enhance catalytic activity, its capacity to affect the photochemical process favorxably has been underexplored by comparison.
Here, we study the impact of surface alloy species and concentration on hot-carrier generation in Ag nanoparticles.
By first-principles simulations, we photoexcite the localized surface plasmon, allow it to dephase, and calculate spatially and energetically resolved hot-carrier distributions.
We show that the presence of non-noble species in the topmost surface layer drastically enhances hot-hole generation at the surface at the expense of hot-hole generation in the bulk, due to the additional d-type states that are introduced to the surface.
The energy of the generated holes can be tuned by choice of the alloyant, with systematic trends across the d-band block.
Already low surface alloy concentrations have a large impact, with a saturation of the enhancement effect typically close to \qty{75}{\%} of a monolayer.
Hot-electron generation at the surface is hindered slightly by alloying but here an judicious choice of the alloy composition allows one to strike a balance between hot electrons and holes.
In this context, it is also important to consider that increasing the alloy concentration broadens the localized surface plasmon resonance, and thus decreases hot-carrier generation overall.
Our work underscores the promise of utilizing multicomponent nanoparticles to achieve enhanced control over plasmonic catalysis, and provides guidelines for how hot-carrier distributions can be tailored by designing the electronic structure of the surface through alloying.
\end{abstract}

\maketitle

\section{Introduction}

Several emerging technologies in light-harvesting \cite{GenAbdBer21}, solar-to-chemical energy conversion \cite{AslRaoCha18, LiCheRic21, DuCTagWel18}, and catalysis \cite{ZhoLouBao21, DuCTagWel20, HouCheXin20, YamKuwMor21} rely on \gls{hc} generation in plasmonic \glspl{np}.
During this process, light is absorbed in \glspl{np}, creating a collective electronic excitation \cite{KreVol95} that decays into highly non-thermal electrons and holes \cite{BroHalNor15, GonMun15, RomHesLis19, Khu19, Khu20, HatMenZhe21, HawSilBer21, BerMusNea15, RosKuiPus17, ZhoSweZha18, AslRaoCha18, RosErhKui20, KumRosKui19, KumRosMar19, RosErhKui20, VilLeiMar22}.
These non-thermal carriers (usually called ``hot'', despite being somewhat of a misnomer \cite{KhuPetEic21}) prove useful when they cross some interface, for example to a molecule or to a semiconductor, and can modify chemical reaction barriers \cite{ZhoSweZha18} or contribute to the photocurrent in photovoltaic devices \cite{GenAbdBer21}.
The collective electronic excitation is called a \gls{lsp} \cite{KreVol95} and is particularly strong in noble metal \glspl{np}, manifesting as a large optical absorption cross sections at visible frequencies \cite{Boh83, LanKasZor07}.

\begin{figure*}
    \centering
    \includegraphics{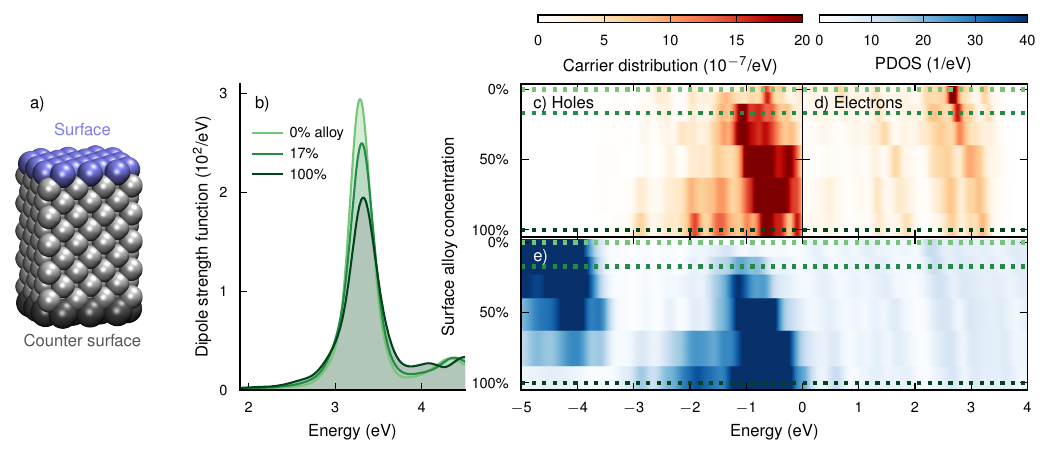}
    \caption{
    \textbf{Distribution of hot carriers at the surface of Ag--Pt \glspl{np} as a function of composition.}
    (a) \Gls{np} geometry studied here, with the top surface marked, which is alloyed and where the carrier distributions are evaluated.
    (b) Absorption spectra of Ag--Pt \glspl{np} with different Pt surface concentrations.
    (c) Hole and (d) electron distributions at the top surface after resonant laser excitation as a function of surface composition.
    (e) \Glsentrylong{pdos} (\glsentryshort{pdos}) of the top surface.
    The dotted lines in (c--e) correspond to the spectra shown in (b).
    }
    \label{fig:surface_alloying}
\end{figure*}

Prototypical plasmonic metals such as Ag \cite{ChrXinLin11, YamKuwMor21}, Au \cite{DuCTagWel18, LiCheRic21, SahYanMas22} or Cu \cite{DuCTagWel20, HouCheXin20} are used due to their outstanding optical properties.
However, recently there has been increased interest in multicomponent \glspl{np}, such as antenna-reactor \cite{ZhoMarFin20, RenYanYan21, JinHerCor23}, core-shell \cite{AslChaLin17} or single-atom alloys \cite{ZhoMarFin20, SorRinRos23}.
This interest is motivated by the fact that typical plasmonic metals (Ag, Au or Cu) have the right optical properties but are poor traditional catalysts.
In fact, using \glspl{np} with a plasmonic core and a catalytic surface alloy, several groups \cite{AslChaLin17, ZhoMarFin20, RenYanYan21} have achieved better photocatalytic rates than with single-component systems.
Several mechanisms can lead to improved reaction rates, and as the processes take place on picosecond or femtosecond scale they can be hard to distinguish.
Assuming that the reaction barrier is lowered by an occupation change in an orbital of the reactant \cite{ZhoSweZha18}, the charge transfer can take place either directly by the \gls{lsp} dephasing into a charge transfer excitation \cite{ChrXinLin11, KhuPetEic21, MaGao19, FojRosKui22}, or indirectly by scattering of a \gls{hc} from the reactive surface of the \gls{np}.
In the latter case, \glspl{hc} need to be generated at the surface (directly through the decay of the \gls{lsp} or through electromagnetic field enhancement, often called plasmon induced resonant energy transfer \cite{CusLiMen12, RenYanYan21, RudKafWat23}) or scattered from \glspl{hc} generated throughout the \gls{np} \cite{EngCraFis20}.
To add further complexity, all processes eventually lead to local heating, which by itself usually increases catalytic activity, and care needs to be taken to disentangle these effects experimentally \cite{DubUnSiv20, Jai20, ZhoSweZha18, SivBarUn19, ZhoSweRob19}.
Theoretical and computational modeling can provide insights into these processes, allowing the rational design of efficient devices \cite{ManLiuKul14, LonPre14, BerMusNea15, BroSunNar16, RomHesLis19, SivUnDub19, RomKahHes20, DubUnSiv20, RosErhKui20, KhuPetEic21, ChaFanWan21, FojRosKui22, JinKahPap22, DubUnSiv22, KluAnt23}.
Strategies for optimizing \gls{hc} generation rates and tailoring \gls{hc} distributions are particularly valuable.

In this work, we study the influence of surface alloying on \gls{hc} generation in plasmonic Ag \glspl{np}.
We find that already a modest surface alloy concentration in core-shell or core-crown configurations can enhance the generation of hot holes, and that the d-band position of the alloy dictates the energy distribution of the holes.
We emphasize that in our \glspl{np}, the shell is photocatalytically active, in contrast to earlier work that considered a photoactive core and a catalytic shell \cite{AslChaLin17, RobZhaSwe17}.
We model plasmon decay and \gls{hc} formation using methods developed in our group \cite{RosKuiPus17, KumRosKui19, KumRosMar19, RosErhKui20} based on \gls{rttddft} \cite{YabBer96}.
We drive our systems with an ultra-short laser pulse, simulate the electron dynamics until the plasmon has decayed, and then analyze the distribution of carriers over the ground state \gls{ks} states.
The dephasing process of the \gls{lsp} into \glspl{hc} has been studied in detail before in Ref.~\citenum{RosErhKui20, FojRosKui22}, which also provide a detailed description of the methodology.

\section{Results and discussion}
\label{sec:results}

\subsection{Tuning the HC distribution via surface composition: Getting the holes to the surface}
\label{subsec:surface-generation}

\begin{figure*}
    \centering
    \includegraphics{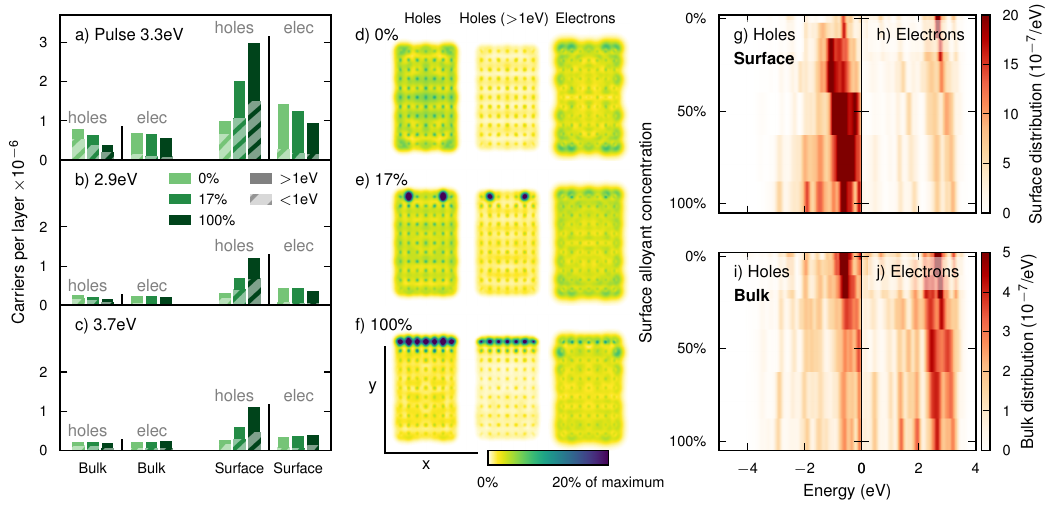}
    \caption{
    \textbf{Hot carrier distribution in Ag--Pt \glspl{np} as a function of pulse energy and location.}
    (a-c) Number of induced carriers in the surface and bulk, respectively after resonant (\qty{3.3}{\electronvolt}) and off-resonant (\num{2.9}/\qty{3.7}{\electronvolt}) laser excitation, for the unalloyed, optimally alloyed (\qty{17}{\%}), and fully alloyed \gls{np}.
    The portion of carriers with an energy of more than \qty{1}{\electronvolt} relative to the Fermi level are shown by solid bars (referred to as ``hot carriers'' in the text).
    The fraction of carriers with an energy below this threshold is indicated by hatched bars.
    (d-f) Visualization of hole, hot hole, and electron densities after resonant laser excitation for unalloyed, optimally alloyed (\qty{17}{\%}), and fully alloyed \glspl{np}.
    The densities have been integrated over the $z$-direction.
    (g) Hole and (h) electron distributions in the surface and (i-j) in the bulk, as a function of surface alloy concentration.
    }
    \label{fig:holes_vacuum}
\end{figure*}

We study the influence of alloying on \gls{hc} formation by considering a few geometrically identical \glspl{np}.
We compare unalloyed Ag with core-crown Ag--Pt \glspl{np} where substitutions are made in the top surface (\autoref{fig:surface_alloying}a; see \nameref{sec:methods} for details).
These \glspl{np} have a \gls{lsp} resonance at \qty{3.3}{\electronvolt} corresponding to excitation along their long axis (\autoref{fig:surface_alloying}b).
The maximum of the \gls{lsp} decreases while the peak broadens with increasing surface alloy concentration, in agreement with experiment \cite{AslChaLin17}.

The surface hole distribution (\autoref{fig:surface_alloying}c, where the ``surface'' is defined in \autoref{fig:surface_alloying}a), following resonant \gls{lsp} excitation (laser pulse $\hbar\omega_\text{pulse} = \qty{3.3}{\electronvolt}$), depends sensitively on the alloy concentration:
The unalloyed \gls{np} has relatively few holes in the surface, in the energy range between $-\hbar\omega_\text{pulse}$ and \qty{0}{\electronvolt}.
Already a modest surface alloy concentration of \qty{17}{\%} greatly increases the hole distribution, in particular between \num{-1.5} and \qty{-0.5}{\electronvolt}.
While the total number of holes (\textit{i.e.}, the integral of the distribution) \emph{increases} with surface alloy concentration, the peak of the distribution also shifts closer to the Fermi level, thus yielding ``colder'' holes.
In contrast, the total number of surface electrons \emph{decreases} as Pt is added to the surface (\autoref{fig:surface_alloying}d) while the distribution shifts to higher energies, corresponding  to ``hotter'' electrons.

The \gls{pdos} in the surface layer (\autoref{fig:surface_alloying}e) indicates which surface states are available.
In the unalloyed \gls{np}, the \gls{pdos} consists of many occupied and unoccupied Ag sp-states above about \qty{-4}{\electronvolt} as well as d-states below approximately \qty{-4}{\electronvolt}.
The latter states do not appear in the hole distribution, as they are further away from the Fermi level than the energy supplied by the resonant laser pulse (\qty{3.3}{\electronvolt}).
As Pt is substituted into the surface, Pt d-states appear between \num{-1.5} and \qty{-0.5}{\electronvolt}, while the number of Ag d-states (around \qty{-4}{\electronvolt}) decreases.
The gradual shift of a Ag-like d-band to a Pt-like d-band with increasing concentration coincides with the increased amount of hole formation after resonant laser excitation.
With increasing Pt concentration, additionally the sp-states shift to higher energies, which is reflected in the electrons becoming ``hotter''.
We note that while for the 269-atom \gls{np} studied here, which is about \qty{2}{\nano\meter} in size, the density of states is discrete, for larger \glspl{np} it would approach a continuum.

\subsection{Why does it work: Localizing the holes at the surface}

\begin{figure}[b]
    \centering
    \includegraphics{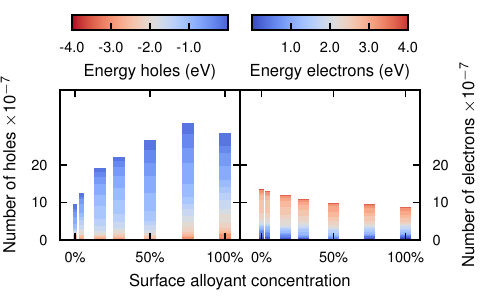}
    \caption{
    \textbf{Number of \glspl{hc} at the surface of the core-crown Ag--Pt \gls{np} for different surface alloy compositions} by energetic range after excitation with a laser at the \gls{lsp} peak (\qty{3.3}{\electronvolt}).
    }
    \label{fig:bar_plot}
\end{figure}

For alloyed \glspl{np}, the increase in \emph{hole carriers} at the surface comes at the expense of holes in the bulk (\autoref{fig:holes_vacuum}a; also see \autoref{fig:bar_plot}).
While the number of holes at the surface doubles (\qty{100}{\%} increase) for a \qty{17}{\%} surface alloy coverage compared to the unalloyed \gls{np}, the number of holes in the bulk is reduced by \qty{20}{\%}.
This also applies to ``hot'' holes, which we here define as hole states with an energy of more than \qty{1}{\electronvolt} below the Fermi energy.
The concentration of the latter is enhanced almost threefold (\qty{183}{\%}) at the surface while their concentration in the bulk is reduced by \qty{2}{\%} compared to the pure Ag \gls{np}.
In the fully alloyed surface, these numbers increase to \qty{196}{\%}/\qty{337}{\%} for all/hot holes in the surface and a reduction of \qty{50}{\%}/\qty{24}{\%} in the bulk.
Alloying thus pulls holes from the bulk to the alloyed surface, while the total number of holes decreases somewhat due to the broader \gls{lsp} resonance (\autoref{fig:surface_alloying}b).

The total number of \emph{excited electrons} is also decreased by alloying, but the decrease is more severe in the surface than in the bulk (\autoref{fig:holes_vacuum}a).
Hence there is a trade-off when alloying between increasing the amount of holes or electrons in the surface, such that at, \textit{e.g.}, \qty{17}{\%} surface alloy concentration one can excite both many electrons and holes (\autoref{fig:bar_plot}).
Similar trends are observed for the hole distribution when exciting the system using an off-resonant pulse (\autoref{fig:holes_vacuum}b-c), while the electron distributions change less systematically.

\begin{figure}
    \centering
    \includegraphics{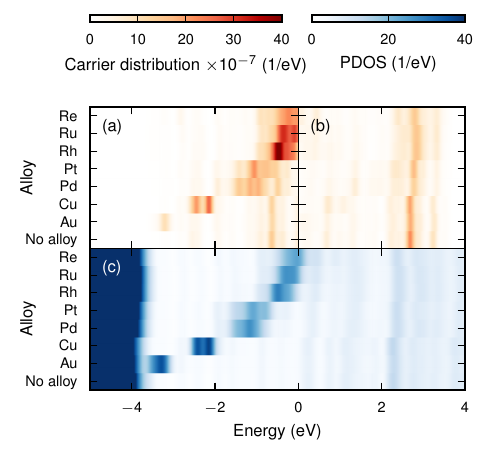}
    \caption{
    \textbf{Variation of hot carrier distribution with alloyant.}
    (a) Hole and (b) electron distributions for Ag-alloy \glspl{np} with a surface composition of \qty{17}{\%} after excitation with a laser at \gls{lsp} peak (\qty{3.3}{\electronvolt}).
    (c) \Gls{pdos} in the top surface layer.
    }
    \label{fig:elements_hcdist}
\end{figure}

Electrons and holes differ qualitatively in how they are affected by alloying because electrons always populate sp-states, which are delocalized in character, and holes may populate the Pt d-type states, which are localized.
In the unalloyed \gls{np}, holes, hot holes, and electrons are fairly evenly distributed over the \gls{np} (\autoref{fig:holes_vacuum}d).
The electrons are only in slight excess at the edges of the \gls{np}, which is due to preferential localization of carriers to undercoordinated surface sites \cite{RosErhKui20}.
In the \glspl{np} with surface alloy concentrations of \qty{17}{\%} and \qty{100}{\%}, a large fraction of holes and hot holes are localized at alloy sites, while the density of electrons remains relatively uniform (\autoref{fig:holes_vacuum}e-f).
The localization of holes at the surface also becomes apparent by comparing the surface hole distribution (\autoref{fig:holes_vacuum}g) to the bulk hole distribution (\autoref{fig:holes_vacuum}i), where the former even shows a non-monotonic variation with composition.
In contrast, electron distributions in the surface (\autoref{fig:holes_vacuum}h) and bulk (\autoref{fig:holes_vacuum}j) are similar to each other because they are delocalized over the entire \gls{np}.
The electron distributions are relatively unaffected by alloying except for an overall shift of states and decrease in intensity.

The number of hole carriers at the surface increases with alloying until it reaches a maximum at \qty{75}{\%} alloy concentration (\autoref{fig:bar_plot}), with the steepest increase at low concentrations.
To understand why the increase is not simply linear with the alloy concentration (even after compensating for the lowered absorption; see \autoref{sfig:normalized_fraction}), as one might expect, the mechanism of carrier formation has to be considered.
Carriers are formed in pairs after plasmon decay, with the energetic difference between electron and hole equal to $\hbar\omega_\text{pulse}$ \cite{FojRosKui22}.
(Note that we do not consider electron-electron or electron-phonon scattering processes in our calculations.)
The pulse frequency and width thus determine which electron hole pairs can form.
However, the probability of electron-hole pair formation depends additionally on the coupling strength of the pair to the \gls{lsp}.
The saturation of hole formation at high concentrations could thus be caused by screening of the d-type holes and their interaction with the \gls{lsp}.

For completeness, we also consider alloying of more than one full surface layer on one side of the \gls{np} (core-crown alloy) as well as layers on both sides (core-shell alloying; \autoref{sfig:layers_sum} and \autoref{sfig:layers_distr}).
As expected, this increases the total amount of holes in the surface (counting all alloyed layers), but there are fewer holes per layer due to the further decreased absorption.

% Most notably, the electron distribution in the alloyed \gls{np} is missing the big peak around \qty{2.5}{\electronvolt}.
% The differences can stem from either differences in the ground state, or in the field due to the \gls{lsp}.
% The distribution is broad, with excited electrons in the range 0 to $\hbar\omega_\text{pulse}$ that stem from occupied states both in the Ag and Pt layers.
% Keep in mind that as the unoccupied states are delocalized, transitions in the entire \gls{np} contribute to the surface \gls{he} distribution.

Finally, we note that we can also generate holes in the d-band of Ag by using pulses that are sufficiently energetic to excite transitions to unoccupied states, \textit{i.e.}, for $\gtrapprox \qty{3.8}{\electronvolt}$).
The resulting hole densities are, however, notably lower than for less energetic pulses and the holes form predominantly in the bulk (\autoref{sfig:heatmap}).
Highly energetic holes in the bulk are usually not the intended outcome, so this is not so relevant for the Ag-core-Pt-shell system.
It could, however, be of interest in, \textit{e.g.}, core-shell structures with an Ag shell and a core lacking a d-band (\textit{e.g.}, Al).

\section{Tuning the \texorpdfstring{\glsentryshort{hc}}{HC} distribution through chemistry: Moving the distribution in energy}
\label{sec:element-alloying}

\begin{figure}
    \centering
    \includegraphics{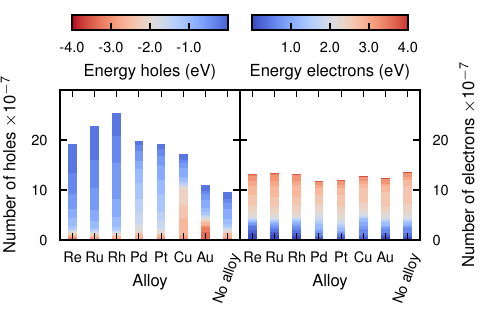}
    \caption{
    Number of hot carriers in the top surface for Ag-alloy \glspl{np} with a surface composition of \qty{17}{\%} after excitation with a laser at the \gls{lsp} peak (\qty{3.3}{\electronvolt}).
    }
    \label{fig:bar_plot_elements}
\end{figure}

It is now instructive to explore the effect of the character of the alloyant on the hot carrier distribution.
To this end, we compare the \gls{hc} distribution for seven different alloyants from the d-block of the periodic table, while keeping the pulse frequency fixed to the \gls{lsp} peak and the surface alloy concentration at \qty{17}{\%} (\autoref{fig:elements_hcdist}a-b and \autoref{fig:bar_plot_elements}).
The \gls{pdos} in the surface layer (\autoref{fig:elements_hcdist}c) shows the d-states of the alloyant shifting closer to the Fermi level as the alloyant is found further to the left in the periodic table, \textit{i.e}, as the number of electrons in the outermost d-shell of the alloyant decreases.

The hole distribution appears to be a mixture of the bare Ag \gls{np} hole distribution (holes between \num{-1} and \qty{-0.5}{\electronvolt}, corresponding to sp-states) and holes in the d-states of the alloyant, which is to be expected, as the surface consists of both Ag and alloyant atoms.

The energetic distribution of electrons is almost independent on the alloyant, as the valence band structure is relatively similar for all considered alloys, and because the unoccupied states are delocalized over the entire \gls{np}.
The variation of the hot carrier distribution with surface concentration is similar as in the case of Pt (\autoref{sfig:bar_plot_elements_coverage}).

\begin{figure}
    \centering
    \includegraphics{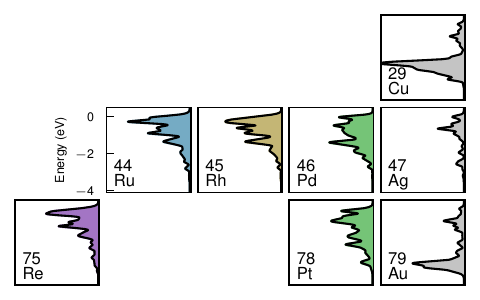}
    \caption{
    Hot hole distribution at the surface of Ag-alloy \glspl{np} with a surface composition of \qty{17}{\%} after excitation with a laser at \gls{lsp} peak (\qty{3.3}{\electronvolt}).
    }
    \label{fig:ptable}
\end{figure}

The working principle of \gls{hc} generation in these \glspl{np} is that a \gls{lsp} is induced in the Ag core by absorbing light, and decays into excited electron-hole pairs.
The electron-hole pairs consist of both intraband sp-sp transitions and interband d-sp transitions, where the former are entirely delocalized and the latter consist of localized holes and delocalized electrons.
The alloyants provide occupied d-states at the surface, allowing holes to form.
By controlling the alloyant concentration and species the hot-hole distribution can thus be tuned (\autoref{fig:ptable}).
The group 10 transition metals Pt and Pd produce holes between \num{-2} and \qty{0}{\electronvolt}.
Moving to the left of the d-block of the periodic table, the d-band shifts closer to the Fermi level, so that the group 9, 8, and 7 elements Rh, Ru and Re generate holes between \num{-1} and \qty{0}{\electronvolt}.

\section{Conclusions and outlook}
\label{sec:conclusions}

We have modeled core-crown \glspl{np} with plasmonic Ag cores and transition metal crowns.
Our results show that core-crown alloying can be effective for enhancing \gls{hc} distributions at the surface of nanostructures.
In particular, we have shown that the type of alloyant influences the energy of the holes in the resulting surface hot-hole distribution (\autoref{fig:ptable}) but has a very minor impact on the surface hot-electron distribution.
The alloyant surface concentration determines the intensity of the \gls{hc} distributions, as the number of induced holes increase with concentration while the number of electrons decreases.

In general, alloying broadens the optical absorption peak, and thus typically reduces the amount of energy absorbed when exciting the system with a laser resonant with the \gls{lsp}.
In part because of the lower absorption, the amount of excited electrons decreases with alloyant concentration.
One should, however, probably also consider the additional effect of the intraband transitions coupling more weakly to the \gls{lsp} when alloying.
The mechanism of such an effect is hard to pinpoint.
Despite the lower total absorption, the amount of holes increases non-linearly with increasing surface alloy concentration, and saturates around \qty{75}{\%}.
The saturation could be explained by the presence of many holes in d-states screening the coupling between the plasmon and the interband transitions.

Based on our results, we would suggest to design as thin shells as possible in core-shell setups.
Alloying less than one full layer is actually preferable as this approach enables the biggest gains in hot-hole enhancement without suffering a big loss in hot-electron generation.
Even if for larger \glspl{np} the optical spectrum is not as significantly affected, there is the presumed effect of screening of the interband transitions.

In our work, the \gls{hc} distributions reach a steady state after a few tens of femtoseconds, because our model does not include decay channels such as reemission, Auger scattering (would require a non-adiabatic \gls{xc}-kernel) or electron-phonon scattering.
For the \glspl{hc} to do any useful work, such as catalyzing a chemical reaction, they would need to transfer across the interface to another system, possibly undergoing scattering processes in the \gls{np}; a process that may take several picoseconds \cite{BroSunNar17}.
From an application point of view, the observable of interest would be the rate of \gls{hc} transfer to the system of interest (or the catalytic rate) that is in competition with the various decay channels.
Modeling these processes is beyond the scope of our work.
Despite studies showing that relaxation times are somewhat dependent on alloyant \cite{MemPalMur20, VilLeiMar22}, we can expect that a higher steady state \gls{hc} distribution (that is barring the decay channels) predicted in our work, corresponds to a higher catalytic rate.

The \gls{hc} distributions depend weakly on the pulse frequency, in the sense that the intensity, but not the shape, changes.
We thus expect our analysis using a narrow Gaussian laser pulse in resonance with the \gls{lsp}, to be also applicable for absorption of solar light.

\section*{Methods}
\label{sec:methods}

\subsection{Structures}

All \glspl{np} that we consider are identical in geometry: cuboids consisting of 7-by-11 atomic layers (1.2-by-\qty{2}{\nano\meter}) of a fcc lattice.
The lattice constant is \qty{4.09}{\angstrom} and we have not relaxed the structures, in order to study the effect of chemistry and not local geometry effects.
We consider core-crown \glspl{np} alloys, where the top surface atoms (\autoref{fig:surface_alloying}a) have been swapped to the alloy species, and core-shell \glspl{np}, where both top and bottom surfaces have been swapped.
Additionally, we consider several-layer swaps and partial layer swaps (\autoref{sfig:geometry}).

\subsection{Computational details}

The open-source \gpaw{}\cite{MorHanJac05, EnkRosMor10} code package was used for all calculations.
\Gls{ks} density functional theory ground state calculations were performed within the projector augmented wave \cite{Blo94} formalism using \gls{lcao} basis sets \cite{LarVanMor09}; the \emph{pvalence} \cite{KuiSakRos15} basis set, which is optimized to represent bound unoccupied states, was used for the metal species.
The PBE \cite{PerBurErn96, PerBurErn97} functional with a Hubbard $+U$ correction \cite{LieAniZaa95} in the form by Dudarev \textit{et al.} \cite{DudBotSav98} was used, with $U$ values \qty{3.5}{\electronvolt} for Ag, \qty{2.5}{\electronvolt} for Au, and \qty{4.5}{\electronvolt} for Cu.
A simulation cell of \qtyproduct{25.6 x 25.6 x 38.4}{\angstrom} was used to represent wave functions, \gls{xc}, and Coulomb potentials, with a grid spacing of \qty{0.2}{\angstrom} for wave functions and \qty{0.1}{\angstrom} for potentials.
The Coulomb potential was represented in numerical form on the grid, with an additional analytic moment correction \cite{CasRubSto03} centered at the \gls{np}.
Fermi-Dirac occupation number smearing with width \qty{0.05}{\electronvolt} was used.
The self-consistent loop was stopped when the integral of the difference between two subsequent densities was less than \qty{1e-12}{}.
Pulay \cite{Pul80}-mixing was used to accelerate the ground state convergence.

The \gls{lcao}-\gls{rttddft} implementation \cite{KuiSakRos15} in \gpaw{} was used for the \gls{rttddft} calculations.
A \deltakick{} strength of $K_z = 10^{-5}$ in atomic units was used.
The time propagation was done in steps of \qty{10}{\atto\second} for a total length of \qty{30}{\femto\second} using the adiabatic PBE+U kernel.
We computed carrier generation for an external electric field corresponding to an ultra-short Gaussian laser pulse
\begin{align}
    \label{eq:laser}
    \mathcal{E}_z(t) = \mathcal{E}_0 \cos(\omega_0 t) \exp(- (t-t_0)^2/\tau_0^2)
\end{align}
of frequency $\omega$, strength $\mathcal{E}_0 =  \qty{51}{\micro\volt\per\angstrom}$, peak time $t_0 = \qty{10}{\femto\second}$, and duration $\tau_0 = \qty{3.0}{\femto\second}$.
Following the methods of Refs.~\citenum{RosErhKui20, FojRosKui22}, the computation of hot carrier generation was carried out by convoluting the first order density response of the \deltakick{}-calculation with the laser pulse, and the hot-carrier distributions were projected according to the atomic layer Voronoi weights.

We computed the total density of states as
\begin{align}
    \sum_k \delta(\varepsilon - \varepsilon_k)
\end{align}
and the \gls{pdos} for the atomic layers as
\begin{align}
    \sum_k \delta(\varepsilon - \varepsilon_k) \int_\text{layer} \left|\phi^{(0)}_k(\vec{r})\right|^2 \dd{\vec{r}},
\end{align}
where $\varepsilon_k$ and $\phi^{(0)}_k(\vec{r})$ are the \gls{ks} eigenvalues and wave functions.
For visualization, the $\delta$-functions in energy were replaced by a Gaussian $(2\pi\sigma^2)^{-1/2} \exp(-\varepsilon^2/2\sigma^2)$ with width $\sigma=\qty{0.05}{\electronvolt}$.

\section*{Data Availability}

The data generated in this study are openly
available via Zenodo at \url{https://doi.org/10.5281/zenodo.10047664}.

\section*{Software used}
The \gpaw{} package \cite{MorHanJac05, EnkRosMor10} with \gls{lcao} basis sets \cite{LarVanMor09} and the \gls{lcao}-\gls{rttddft} implementation \cite{KuiSakRos15} was used for the \gls{rttddft} calculations.
The PBE \cite{PerBurErn96,PerBurErn97} \gls{xc}-functional, utilizing the Libxc\cite{LehSteOli18} library, were used in \gpaw{}.
The \textsc{ase} library \cite{LarMorBlo17} was used for constructing and manipulating atomic structures.
The NumPy \cite{HarMilvan20}, SciPy \cite{VirGomOli20}, and Matplotlib \cite{Hun07} Python packages and the VMD software \cite{HumDalSch96, Sto98} were used for processing and plotting data.

\section*{Acknowledgements}
We acknowledge funding from the Knut and Alice Wallenberg foundation (Grant No.~2019.0140; J.F. and P.E.), the Swedish Research Council (No.~2020-04935; J.F. and P.E.), and the Academy of Finland (No.~332429; T.P.R.).
P.V.K. acknowledges the Scientia Fellowship scheme at The University of New South Wales and the Australian Research Council for financial support through the Discovery Early Career Researcher Award (DE210101259).
The computations were enabled by resources provided by the National Academic Infrastructure for Supercomputing in Sweden (NAISS) at NSC, PDC and C3SE partially funded by the Swedish Research Council through grant agreement no. 2022-06725.

\section*{Supporting information}

\begin{itemize}
\item Figures S1 to S7 displaying supporting data.
\end{itemize}

\end{document}